\documentclass[a4paper,11pt]{article}
\usepackage{graphicx}
\usepackage{amsfonts}
\usepackage{amsmath}
\usepackage{hyperref}

\title{Provably insecure group authentication: Not all security proofs are what they claim to be (Corrected version)}
\author{Chris J. Mitchell\\Information Security Group, Royal Holloway, University of London\\
ORCiD: 0000-0002-6118-0055\\
{\tt me@chrismitchell.net}\\
\url{www.chrismitchell.net}\\
}

\date{9th June 2021}
\newcommand{\qed}{\nobreak \ifvmode \relax \else
      \ifdim\lastskip<1.5em \hskip-\lastskip
      \hskip1.5em plus0em minus0.5em \fi \nobreak
      \vrule height0.75em width0.5em depth0.25em\fi}
\begin{document}

\maketitle

\begin{abstract}
A paper presented at the ICICS 2019 conference describes what is claimed to be a `provably secure
group authentication [protocol] in the asynchronous communication model'.  We show here that this
is far from being the case, as the protocol is subject to serious attacks. To try to explain this
troubling case, an earlier (2013) scheme on which the ICICS 2019 protocol is based was also
examined and found to possess even more severe flaws --- this latter scheme was previously known to
be subject to attack, but not in quite as fundamental a way as is shown here. Examination of the
security theorems provided in both the 2013 and 2019 papers reveals that in neither case are they
exactly what they seem to be at first sight; the issues raised by this are also briefly discussed.
\end{abstract}

\section{Introduction} \label{section-intro}

A paper presented at ICICS 2019 \cite{Xia20} describes a protocol designed to enable members of a
group to authenticate one another in a group-wise fashion.  The paper also presents a formal
security model for such `group authentication' schemes, and provides proofs of security for the
protocol.  Unfortunately, as we describe in this paper, the protocol is completely insecure,
allowing an outsider to masquerade as any group member and set up contradictory views of group
authentication membership within a set of participating entities.

The fact that a fundamental flaw exists in a provably secure scheme is perhaps surprising. However,
as we discuss in greater detail below, examination of the main theorems reveals that they do not
establish that the protocols are secure in any practical sense.

It turns out that the ICICS 2019 scheme is related to a scheme presented in 2013 \cite{Harn13} by
one of the authors of the 2019 paper. As we discuss below, this earlier scheme is also completely
insecure. Since the 2013 scheme is slightly simpler than the 2019 scheme, we present it and its
flaws first, before doing the same for the 2019 scheme.  We observe that the 2013 scheme has been
cryptanalysed previously by Ahmadian and Jamshidpur \cite{Ahmadian18}, although the attack we
describe here is very much simpler than the previously published attack.

The remainder of the paper is structured as follows.  In \S\ref{section-2013-scheme} the 2013
protocol is described, and the goals of, and security claims for, the protocol are summarised. This
leads naturally to \S\ref{section-2013-analysis} in which it is shown that the claimed security
properties do not hold by describing a very simple attack; the `proofs' of the failed theorems are
also examined to see why an apparently provably secure scheme is fundamentally flawed.
\S\ref{section-2020-scheme} presents the ICICS 2019 protocol, together with a summary of its design
goals and security claims. This is followed by \S\ref{section-2020-analysis}, where we show why it
also possesses fundamental flaws; again the security theorems are examined. Conclusions are drawn
in \S\ref{section-conclusions}.

It is important to observe that this is a major revision of the version of the paper published in
the Proceedings of ICSP 2020 \cite{Mitchell21}.  The description of the ICICS 2019 scheme in
Section~\ref{sub-2020-operation} has been corrected, and the method of attack in
Section~\ref{section-2020-analysis} has correspondingly changed.

\section{The 2013 Harn Scheme}  \label{section-2013-scheme}

\subsection{Goals of the Scheme}  \label{sub-2013-goals}

In the context of the schemes considered in this paper, a \emph{group authentication protocol} is
one in which `each user acts both roles of the prover and the verifier, and all users in the group
are authenticated at once' \cite{Xia20}.  The primary goal of such protocols is speed and
efficiency, and not privacy (since all users in such a protocol are identified to each other).  As
discussed, for example, by Yang et al.\ \cite{Yang14}, this contrasts with the use of the same or
similar terms elsewhere in the literature, where protocols are considered which allow an entity to
authenticate to another party as a member of a group, without revealing his or her identity.

The main goal for a group authentication protocol as considered here is to enable all members of a
defined group to be given assurance, through executing the protocol, that the specified members are
all present and actively involved in the protocol, and that no other parties are involved.  A
review of recent work on the design of such protocols can be found in \S 1.1 of Xia et al.\
\cite{Xia20}.

Unfortunately, the precise threat model for which the protocol was designed is not clear from the
2013 paper, \cite{Harn13}.  References are made to both insider and outsider attacks, i.e.\ the
protocol is intended to be secure against both of these classes of attack. However, no reference is
made to the trust assumptions for the broadcast channel used for communication between the parties.
However, it is standard practice when analysing authentication protocols to assume that an attacker
can manipulate the communications channel, including to intercept, delete, insert and modify
messages (see, for example, Boyd et al. \cite{Boyd20}, \S1.5.1). We therefore assume this in our
analysis of the scheme. Indeed, it is hard to imagine a practical situation where it would not be
possible for a determined attacker to modify messages; certainly there are many real-world examples
of message manipulation attacks on the broadcast channels used in mobile telephony --- see, for
example, the rich literature on IMSI catcher attacks \cite{Dabrowski14,Khan17,Mjolnes17}.

\subsection{Operation}  \label{sub-2013-operation}

Harn \cite{Harn13} actually presents three distinct protocols.  The first, the `basic scheme' is
intended to demonstrate the main ideas; however, it requires information to be divulged
simultaneously by all parties and hence would not be secure in practice. The second and third
schemes are elaborations of the basic idea designed to allow for asynchronous information release.
In the second scheme participant credentials can only be used once, whereas the third scheme allows
multiple uses of credentials. However, since the second and third schemes are very similar in
operation, for simplicity we focus here on the second scheme.

\subsubsection{Initialisation}

This scheme, like all the schemes in both papers, involves a \emph{Group Manager} (GM) trusted by
all participants, which pre-equips all participants with credentials used to perform the group
authentication process.  We suppose that there are $n$ participants
$\mathcal{U}=\{U_1,U_2,\ldots,U_n\}$.

To initialise the protocol, the GM performs the following steps.
\begin{itemize}
\item The GM chooses parameters $t$ and $k$, where $t$ determines the resistance of the scheme
    to insider adversaries --- that is, the scheme is designed to be secure so long as at most
    $t-1$ insiders collaborate.  No explicit guidance on the choice of $k$ is given except that
    it must satisfy $kt>n-1$, and hence here we assume $k=\lceil n/t\rceil$.
\item The GM chooses a large prime $p$.  All calculations are performed in
    $\mbox{GF}(p)=\mathbb{Z}_p$.
\item The GM chooses a cryptographic hash function $H$ with domain $\mathbb{Z}_p$.
\item The GM chooses a secret $s\in\mathbb{Z}_p$, and computes $H(s)$.
\item The GM selects a set of $k$ polynomials $\{f_1(x),f_2(x),\ldots,f_k(x)\}$ over
    $\mathbb{Z}_p$ of degree $t-1$, where the coefficients are chosen uniformly at random from
    $\mathbb{Z}_p$.
\item The GM selects two sets of $k$ integers $\{w_1,w_2,\ldots,w_k\}$ and
    $\{d_1,d_2,\ldots,d_k\}$ with the property that
    \[ s = \sum_{j=1}^k d_jf_j(w_j), \]
where the values $\{w_1,w_2,\ldots,w_k\}$ are all distinct.
\item The GM computes a set of $k$ tokens $\{f_1(x_i),f_2(x_i),\ldots,f_k(x_i)\}$ for each
    participant $U_i$ ($1\leq i\leq n$), where $x_i\in\mathbb{Z}_p$ is a unique identifier for
    $U_i$.
\item Using an out-of-band secure channel, the GM equips participant $U_i$ ($1\leq i\leq n$)
    with $t$, $k$, $p$, $H$, the identifiers $\{x_1,x_2,\ldots,x_n\}$, the integers
    $\{w_1,w_2,\ldots,w_k\}$ and $\{d_1,d_2,\ldots,d_k\}$, $H(s)$, and the participant's
    collection of $k$ secret tokens $\{f_1(x_i),f_2(x_i),\ldots,f_k(x_i)\}$.
\end{itemize}

\subsubsection{Group Authentication}  \label{subsub-2013-authentication}

We now suppose that some subset $\mathcal{U}'\subseteq\mathcal{U}$ of the participants (where
$|\mathcal{U}'|=m\leq n$) wish to authenticate each other in  a group-wise fashion.  Suppose
$\mathcal{U}'=\{U_{z_1},U_{z_2},\ldots,U_{z_m}\}$.  We suppose every participant in $\mathcal{U}'$
is aware of the membership of $\mathcal{U}'$.  Each participant $u_{z_i}\in\mathcal{U}'$ now
proceeds as follows.

\begin{itemize}
\item Compute
\[ c_{z_i} = \sum_{j=1}^k d_jf_j(x_{z_i}) \prod_{ \substack{r=1\\r\neq i} }^m\frac{(w_j-x_{z_r})}{(x_{z_i}-x_{z_r})} \]
\item Broadcast $c_{z_i}$ to all members of $\mathcal{U}'$.
\item Once all the values $\{c_{z_1},c_{z_2},\ldots,c_{z_m}\}$ have been received, compute
\[ s' = \sum_{r=1}^m c_r.\]
\item If $H(s')=H(s)$ then the protocol succeeds, i.e.\ all users have been successfully
    authenticated.
\end{itemize}

Note that the protocol can only be executed once per initialisation, as the secret $s$ is revealed
to anyone receiving the messages sent on the broadcast channel. The third scheme removes this
limitation.

\subsection{Security Claims}  \label{sub-2013-claims}

A number of claims are made with respect to the security properties of the protocol.  In particular
the \emph{security} property is claimed, namely that any outside adversary cannot impersonate
\ldots~a member \ldots~after knowing at most $n-1$ values from other members'.  The meaning of
impersonation in this context is not clear, but we assume that this means that, following
completion of the protocol, legitimate participants cannot end up with differing beliefs about who
are the participants in a group authentication.  Sadly, as we show below, this property does not
hold.

\section{Analysis of the 2013 Scheme}  \label{section-2013-analysis}

\subsection{Previous Results}

As noted in \S\ref{section-intro}, this scheme has previously been cryptanalysed by Ahmadian and
Jamshidpour \cite{Ahmadian18}.  Their approach involves performing computations using broadcast
values intercepted during protocol execution, and requires certain conditions to be satisfied to
succeed.  The attack we describe below is almost trivially simple, and works regardless of group
size.

\subsection{Preliminary Observation}  \label{sub-2013-observations}

The attack we propose below relies on a very simple fact.  From the description in
\S\ref{subsub-2013-authentication} it should be clear that participant $U_{z_i}$ will accept that
the group authentication has succeeded if and only if the sum of the $m-1$ received values
$c_{z_j}$ ($j\not=i$) and the value $c_{z_i}$ it computed is equal to $s$.  That is, the
correctness of individual $c_{z_j}$ values is not checked.

\subsection{An Outsider Impersonation Attack}  \label{sub-2013-attack}

Suppose an (insider) adversary controls the broadcast channel with respect to `victim' participant
$U_{z_i}$, i.e.\ the adversary can (a) prevent messages sent by other legitimate participants from
reaching $U_{z_i}$, and (b) send messages to $U_{z_i}$ on this channel that appear to have come
from other legitimate participants.  Since the protocol makes no assumptions about the
trustworthiness of the communications channels (see~\S\ref{sub-2013-goals}), this assumption is
legitimate (indeed, if the broadcast channel was completely trustworthy, then the security protocol
would not be needed).

The adversary does two things.  Firstly it legitimately engages in the protocol with an arbitrary
subset $\mathcal{U}''$ of the legitimate participants, where $U_{z_i}\not\in\mathcal{U}''$.  As a
result of completing this protocol, the adversary now knows $s$.  During execution of the protocol,
the adversary prevents any of the broadcast messages reaching $U_{z_i}$. The adversary now engages
with the `victim' participant $U_{z_i}$, suggesting that a group authentication is to be performed
by the members of an arbitrary set of participants $\mathcal{U}'\subseteq\mathcal{U}$, where
$U_{z_i}\in\mathcal{U}'$ and $|\mathcal{U}'|=m$, say. This may involve sending `fake' messages to
$U_{z_i}$ that apparently originate from the other members of $\mathcal{U}'$.

The adversary now chooses values $c_{z_j}$ ($j\not=i$) for $U_{z_j}\in\mathcal{U}'$, and starts
sending them to $U_{z_i}$ as if they come from the members of $\mathcal{U}'$\@. The only condition
the values must satisfy is that they sum to $s-c_{z_i}$. Of course, this means that the adversary
cannot send all $m-1$ values to $U_{z_i}$ until $c_{z_i}$ is sent by $U_{z_i}$, but the protocol is
meant to be used `asynchronously', i.e.\ where not all participants send their messages at the same
time.  It should be immediately obvious that $U_{z_i}$ will accept the success of the protocol,
although clearly the group authentication that $U_{z_i}$ believes has occurred has not actually
occurred.

Note that the third scheme in the 2013 paper \cite{Harn13} suffers from a precisely analogous
attack.

\subsection{What About the Security Theorems?}

The fact that the protocol is so fundamentally flawed is perhaps surprising given Theorem 2,
\cite{Harn13}, which asserts that the scheme `has the properties of the $t$-secure $m$-user
$n$-group authentication scheme \ldots~if $kt>n-1$'.  This appears to contradict the simple attack
we have just described. The answer is simple --- the `proof' of Theorem 2 only attempts to show
that an adversary cannot forge legitimate values $c_{z_j}$, but the attack does not require this.
Thus it is clear that the `proof' is making unwarranted assumptions about how an attack might be
launched, and as such Theorem 2 is demonstrably not a theorem at all.

To be fair, this shortcoming was already noted by Xia et al.\ \cite{Xia20}, who observe that the
security properties of the 2013 scheme `are only justified by heuristic arguments rather than
formal security proofs'.  Unfortunately, despite a much more formal approach, we show below that
the Xia et al.\ scheme is also completely insecure, and that the threat model underlying the
security arguments is not adequate to reflect real-world attacks.

\section{The Xia-Harn-Yang-Zhang-Mu-Susilo-Meng Scheme}  \label{section-2020-scheme}

\subsection{Goals of Scheme}  \label{sub-2020-goals}

The second protocol we consider here, \cite{Xia20}, is also an example of a group authentication
protocol in the sense given in \S\ref{sub-2013-goals}.  Xia et al.\ \cite{Xia20} go much further
than much of the prior art in attempting to formalise the goals and security model for a group
authentication scheme.  However, even here the specific objectives of such a protocol are left a
little vague.  The following statement is the closest to a formal definition.
\begin{quote}
In general, a group authentication scheme works as follows. The group
manager (GM) generates a number of credentials, and sends each of these credentials to a user in
the group. In the authentication stage, every participating user uses her credential to compute a
token and broadcasts it. Subsequently, every user can use the revealed information to verify
whether all users are belonging to the same group.
\end{quote}

However, as was the case with the 2013 paper analysed above, no explicit references are made to the
trust assumptions applying to the channel used for communications.  As a result, when analysing the
protocol below we make the same (standard) assumptions about this channel as were made for the 2013
protocol, namely that messages are subject to interception, insertion, deletion and/or
modification. As stated in~\S\ref{sub-2013-goals}, it is hard to imagine a real-world deployment
scenario where this would not be possible. However, as we discuss below, the security \emph{proof}
implicitly assumes that the attackers are restricted to being passive interceptors (`honest but
curious'), which is why it is possible to construct a security proof for a protocol that under
reasonable real-world assumptions is subject to a fundamental attack.

\subsection{Operation}  \label{sub-2020-operation}

As is the case for the 2013 protocol, the scheme can be divided into two phases: initialisation,
when the GM equips each participant with the credentials needed to perform group authentication,
and the group authentication phase where a subset of the participants simultaneously authenticate
each other as a group.

\subsubsection{Initialisation}

Again as before we suppose that there are $n$ participants $\mathcal{U}=\{U_1,U_2,\ldots,U_n\}$.
To initialise the protocol, the GM performs the following steps.
\begin{itemize}
\item The GM chooses parameters $t$ and $\ell$, where the scheme is designed to be secure as
    long as at most $t-1$ insiders collaborate, and $\ell$ determines the number of group
    authentication sessions that can be performed before new credentials need to be issued.
\item The GM chooses a cyclic group $G$ (expressed multiplicatively) with order a large prime
    $q$, and randomly selects $g_1,g_2,\ldots,g_\ell$ to be $\ell$ independent generators of
    $G$.
\item The GM chooses a cryptographic hash function $H$ with domain $G$.
\item The GM chooses a secret $s\in\mathbb{Z}_q$, and computes the $\ell$ values $H((g_i)^s)$,
    $1\leq i\leq\ell$.
\item The GM randomly selects a polynomial $f(x)=\sum_{i=0}^{t-1}a_ix^i$ over $\mathbb{Z}_q$ of
    degree $t-1$, where $a_0=s$.
\item The GM computes a credential $s_i=f(x_i)$ for each participant $U_i$ ($1\leq i\leq n$),
    where $x_i\in\mathbb{Z}_p$ is a unique identifier for $U_i$.
\item Using an out-of-band secure channel, the GM equips participant $U_i$ ($1\leq i\leq n$)
    with $t$, $G$, $q$, $H$, the identifiers $\{x_1,x_2,\ldots,x_n\}$, the generators
    $\{g_1,g_2,\ldots,g_\ell\}$, the hash codes
    $\{H((g_1)^s),H((g_2)^s),\ldots,H((g_\ell)^s)\}$, and the participant's own secret
    credential $s_i(=f(x_i))$.
\end{itemize}

\subsubsection{Group Authentication}  \label{subsub-2020-authentication}

Just as in the 2013 scheme, we now suppose that some subset $\mathcal{U}'\subseteq\mathcal{U}$ of
the participants (where $|\mathcal{U}'|=m\leq n$) wish to authenticate each other in a group-wise
fashion.  Suppose $\mathcal{U}'=\{U_{z_1},U_{z_2},\ldots,U_{z_m}\}$.  We suppose every participant
in $\mathcal{U}'$ is aware of the membership of $\mathcal{U}'$.  We further suppose that the set of
participants has reached session number $\sigma$ in the period of use of a particular credential
set, where $1\leq\sigma\leq\ell$. Note that each session must be conducted using a new value of
$\sigma$, and $\sigma$ determines which generator $g_\sigma$ from the set of generators will be
used in this particular protocol instance.

Each participant $u_{z_i}\in\mathcal{U}'$ proceeds as follows.

\begin{itemize}
\item Choose $u_{z_i}\in\mathbb{Z}_q$ uniformly at random, and broadcast $(g_\sigma)^{u_{z_i}}$
    to all other participants.
\item Once the set of values
    $\{(g_\sigma)^{u_{z_1}},(g_\sigma)^{u_{z_2}},\ldots,(g_\sigma)^{u_{z_m}}\}$ has been
    received, compute:
\[ \gamma_i = \prod_{\stackrel{{j\in\{1,2,\ldots,m\}}}{z_j<z_i}} (g_\sigma)^{u_{z_j}}
              \prod_{\stackrel{{j\in\{1,2,\ldots,m\}}}{z_j>z_i}} (g_\sigma)^{-u_{z_j}}, \]
\[ L_i = \prod_{\stackrel{{j\in\{1,2,\ldots,m\}}}{z_j\not=z_i}} \frac{x_{z_j}}{x_{z_j}-x_{z_i}}, \]
and
\[ c_{z_i} = (g_\sigma)^{s_{z_i}L_i}(\gamma_i)^{u_{z_i}}. \]
\item Broadcast $c_{z_i}$ to all members of $\mathcal{U}'$.
\item Once all the values $\{c_{z_1},c_{z_2},\ldots,c_{z_m}\}$ have been received, compute
\[ \prod_{r=1}^m c_{z_r}.\]
\item If $H(\prod_{r=1}^m c_{z_r})=H((g_\sigma)^s)$ then the protocol succeeds, i.e.\ all users
    have been successfully authenticated.
\end{itemize}

\subsection{Security Claims}  \label{sub-2020-claims}

We first observe that Xia et al.\ \cite{Xia20} make the following statement about the assumed
properties of the broadcast channel.
\begin{quote}
Note that the broadcast channel is only assumed to be asynchronous, such that messages sent from
the uncorrupted users to the corrupted ones can be delivered relatively fast, in which case, the
adversary can wait for the messages of the uncorrupted users to arrive, then decide on her
computation and communication, and still get her messages delivered to the honest users on time.
\end{quote}
The security model of Xia et al.\ \cite{Xia20} gives two relevant properties.

The \emph{No forgery} property is as follows.
\begin{quote}
The inside adversary $\mathcal{A}_I$ cannot pass the group authentication by herself.  \ldots  It
is required that $\mathcal{A}_I$ still cannot pass the group authentication by herself in a new
session.
\end{quote}
The \emph{No impersonation} property includes the following statement.
\begin{quote}
The outside adversary $A_O$ cannot impersonate a group member without being detected, even if $A_O$
computes her token after seeing all other users' tokens in the asynchronous networks.
\end{quote}

\section{Analysis of the ICICS 2019 Scheme}  \label{section-2020-analysis}

\subsection{Preliminary Observation}  \label{sub-2020-observations}

The final step of the group authentication phase involves checking that the hash of the product of
the $c_{z_i}$ values equals a hash code provided to all participants by the GM.  That is, the
product of the $c_{z_i}$ values is equal to $(g_\sigma)^s$. So the value used to confirm that group
authentication is successful is fixed for instance $\sigma$ of the protocol, and is thus
independent of the composition of the subset of participants who are authenticating each other.
This suggests the attack we describe immediately below.

\subsection{An Outsider Impersonation Attack}  \label{sub-2020-attack}

The above observation leads to a very simple and powerful attack, enabling impersonation of any set
of participants.  The attack scenario is very similar to that described in \S\ref{sub-2013-attack}.
We suppose an (outsider) adversary controls the broadcast channel with respect to `victim'
participant $U_{z_i}$, i.e.\ the adversary can (a) prevent messages sent by other legitimate
participants from reaching $U_{z_i}$, and (b) send messages to $U_{z_i}$ on this channel that
appear to have come from other legitimate participants.  Finally we assume that it is `time' for a
session using the group generator $g_\sigma$.

In the first stage of the attack we suppose that the (outsider) adversary observes a group of
participants $\mathcal{U}''\subseteq\mathcal{U}$ (where $U_{z_i}\not\in\mathcal{U}''$) engaging in
the protocol. The adversary:
\begin{itemize}
\item intercepts all the $c_{z_i}$ values that are broadcast and as a result learns their
    product, which equals $(g_\sigma)^s$;
\item prevents any of the messages reaching $U_{z_i}$ --- who is in any event not a member of
    the group $\mathcal{U}''$.
\end{itemize}

In the second stage of the attack we suppose that the adversary persuades the victim participant
$U_{z_i}$ that it is being invited to join a group of participants
$\mathcal{U}'\subseteq\mathcal{U}$, e.g.\ by sending `fake' messages from members of $\mathcal{U}'$
to $U_{z_i}$.  The adversary chooses arbitrary values $u_{z_j}$ for every
$U_{z_j}\in\mathcal{U}'-\{U_{z_i}\}$, and for every such value sends $(g_\sigma)^{u_{z_j}}$ to
$U_{z_i}$ as if it comes from $U_{z_j}$. Once $U_{z_i}$ has received all these values, it will
compute its value $c_{z_i}$ which it broadcasts (and is intercepted by the adversary).

The adversary now chooses a set of values $c_{z_j}$ for every $U_{z_j}\in\mathcal{U}'-\{U_{z_i}\}$
with the property that
\[ \prod_{j: U_{z_j}\in\mathcal{U}'} c_{z_j}=(g_\sigma)^s. \]
This is easy to achieve, e.g.\ by choosing all but one of the values at random and solving for the
remaining value using the value $c_{z_i}$ sent by $U_{z_i}$ and the value $(g_\sigma)^s$ obtained
from the first stage of the attack.

The adversary now sends these values to the victim participant $U_{z_i}$ as if they come from the
appropriate participants. Since the product of these values is `correct', the victim will falsely
believe that it is part of a group authentication with a set of participants of whom none believe
they are being authenticated to the victim. Note that the adversary does not need to be a member of
the impersonated group, i.e.\ it can be an outsider.

\subsection{Other Possible Attack Scenarios}  \label{sub-2020-generalisations}

There are many other scenarios in which the observation in \S\ref{sub-2020-observations} could be
used to launch an attack on the protocol.  For example, the two stages of the attack described
above could be run simultaneously --- in essence making up a single instance of the protocol. The
only point at which the information gained from the first stage of the attack is needed in the
second stage is right at the end, and the adversary can wait for the first stage instance to
complete before sending the final values to the victim.

Alternatively, if an attacker could control the broadcast network with respect to two victims, a
range of conflicting beliefs about who has been authenticated to whom could be established.  That
is, once an attacker has observed a participant $U_{z_j}$ output a value $c_{z_j}$, this can be
used to impersonate $U_{z_j}$ in any group the attacker chooses (assuming control over the
broadcast channel).

\subsection{What About the Proof of Security?}

It would seem that this attack breaks the `no forgery' and `no impersonation' properties given in
\S\ref{sub-2020-claims}.  The attacker can be an outsider in both stages of the attack. Certainly
the scenario of the attack is not an unreasonable one for any such protocol.

We observe that Theorem 4 \cite{Xia20} states that `The proposed group authentication scheme
satisfies the no impersonation property, assuming that $H$ is a pre-image resistant hash function
and the DDH assumption holds in $G$'.  The fact that the theorem holds appears to be an artefact of
the fact that the security model does not properly capture insider attacks.

We also observe that the proof of Theorem 4 only deals with the `honest but curious' case, where
all participants are assumed to follow the protocol correctly. The sort of manipulation of messages
and beliefs involved in the attack do not appear to be covered by the proof.  That is, while the
mathematics may be correct, the result does not establish that the protocol would actually be
secure in a real-world deployment (which, of course, it is not).

Indeed, this is partly admitted by Xia et al.\ \cite{Xia20}.  In the concluding section of their
paper, it is stated that `There are two distinct approaches to defining security for cryptographic
protocols: simulation proof and reduction proof. The former is more intuitive because it models
security of the targeted problem via an ideally trusted third party. However, the definitions will
become complicated once all details are filled in. In contrast, the reduction proof yields
definitions that are simpler to describe and easier to work with. However, the adequacy for
modelling the problem is less clear. In this paper, we followed the latter approach, and it is
still open how to provide formal security treatment for group authentication using the simulation
proof.'

\section{Conclusions}  \label{section-conclusions}

We have examined two different group authentication protocols, and found that both possess
fundamental flaws.  Clearly this means that neither of them should be used in practice. Fortunately
there are many well-established and relatively efficient means of performing authentication ---
see, for example, Boyd et al.\ \cite{Boyd20}.

The fundamental flaws in the protocols exist despite the fact that in both cases theorems are
provided asserting their security. Indeed, in the more recent case, the theorems are given within
the context of a formal security model.  This is clearly worrying --- modern cryptography takes as
a fundamental tenet that `proofs of security' are necessary, but clearly they are not of much value
if the proofs are false.

Of course, mistakes in proofs are commonplace, but in these cases the issue is clearly not just a
mistake.  In the earlier paper there is no formal security model, and the theorems are simply
heuristic arguments.  Even in the more recent paper, where the results may well be valid, the
authors themselves admit that the security model used is not sufficient to establish security other
than in a case where the attackers are restricted to behaving in an `honest' fashion. This clearly
suggests that reviewers need the time to carefully review proofs (and the precise details of claims
of security) for adequacy. This flies in the face of the modern obsession with speedy publication,
both for conferences and many journals (e.g.\ \emph{IEEE Access} which allows referees only a week
to complete a review). Perhaps we, as the research community, need to think more carefully about
finding ways to allow reviewers the time and space to write carefully considered and detailed
reviews.


\end{document}